\documentclass[aps,prl,twocolumn,groupedaddress,showpacs]{revtex4}

\usepackage{graphicx}
\begin{document}
%\preprint{}

\title{Andreev Current-Induced Dissipation in a Hybrid Superconducting Tunnel Junction}

\author{Sukumar Rajauria$^{1}$, P. Gandit$^{1}$, T. Fournier$^{1}$, F.W.J. Hekking$^{2}$, B. Pannetier$^{1}$ and H. Courtois$^{1,3}$}
%\thanks{}
\affiliation{$^{1}$Institut N\' eel, CNRS and Universit\'e Joseph Fourier, 25 Avenue des Martyrs, BP 166, 38042 Grenoble, France.
$^{2}$LPMMC, Universit\'e Joseph Fourier and CNRS, 25 Avenue des Martyrs, BP 166, 38042 Grenoble, France.
$^{3}$Institut Universitaire de France.}


\date{\today}

\begin{abstract}
We have studied hybrid superconducting micro-coolers made of a double Superconductor-Insulator-Normal metal tunnel junction. Under subgap conditions, the Andreev current is found to dominate the single-particle tunnel current. We show that the Andreev current introduces additional dissipation in the normal metal equivalent to Joule heating. By analyzing quantitatively the heat balance in the system, we provide a full description of the evolution of the electronic temperature with the voltage. The dissipation induced by the Andreev current is found to dominate the quasiparticle tunneling-based cooling over a large bias range.
\end{abstract}

%\keywords{}

\maketitle

In a tunnel junction between a normal metal (N) and a superconductor (S), the charge transfer occurs mainly through two different mechanisms. The tunneling of a single quasi-particle is possible for electrons or holes with an energy $E$  (compared to the Fermi level $E_F$) larger than the superconductor gap $\Delta$. At low energy, the charge transfer occurs through the Andreev reflection \cite{Andreev,SaintJames}. In the normal metal, an electron (a hole) impinging on the superconducting interface is reflected as a hole (an electron), enabling the transfer of a Cooper pair into (out of) the superconductor. As the energies of the involved electron and hole are located symmetrically around $E_F$, the Andreev reflection does not carry heat through the interface at zero bias. The probability for an incident quasi-particle to follow an Andreev reflection, a specular reflection or a tunnel transfer is given in the ballistic regime (no disorder) by the BTK theory \cite{BTK}. For a N-I-S tunnel junction with an insulator (I) of intermediate or low transparency, the Andreev reflection probability is predicted to be vanishingly small. Taking into account the quasi-particles confinement in the vicinity of the interface, this is no longer true. This confinement can be induced by the disorder or the presence of a second barrier in the normal metal. A single quasiparticle then experiences several collisions with the interface \cite{PRL-VanWees,PRL-Hekking}. The actual Andreev reflection transmission coefficient corresponds to the coherent addition of many individual transmission probabilities. Therefore, the Andreev sub-gap current significantly exceeds the ballistic case prediction \cite{PRL-Kastalsky,PRB-Quirion} and can be modulated by a magnetic flux \cite{PRL-Pothier}.

A quasi-particle current in a N-I-S junction indeed carries both a charge current and a heat current. With a voltage bias smaller than the gap $\Delta/e$, the tunnel current is selectively made out of high-energy electrons (or holes); this cools the electronic population of the normal metal \cite{RMP-Giazotto}. In this way, (S-I-N-I-S) micro-coolers based on a double tunnel junction provide a significant temperature reduction which reaches an optimum at a voltage bias just below the gap. At a very low temperature, the thermal transport in such N-I-S tunnel junctions appears to be still little understood. For instance, an apparent reversal of the normal metal temperature evolution was observed in various experiments \cite{APL-Fisher,PRL-Pekola} and related to a non-BCS density of states of the superconductor \cite{PRL-Pekola}. A clear understanding of this behavior is still missing.

In this letter, we describe an experimental study of the heat transport in a S-I-N-I-S junction, focusing on the very low temperature regime. We show that the phase-coherent Andreev current introduces a significant dissipation in the normal metal. We provide a fully quantitative analysis of the heat transfer in the system which shows that although the Andreev current is a small effect in terms of charge current, the heat it creates has a dominating influence on the heat balance.

\begin{figure}[t]
\begin{center}
\includegraphics[width=8.5 cm]{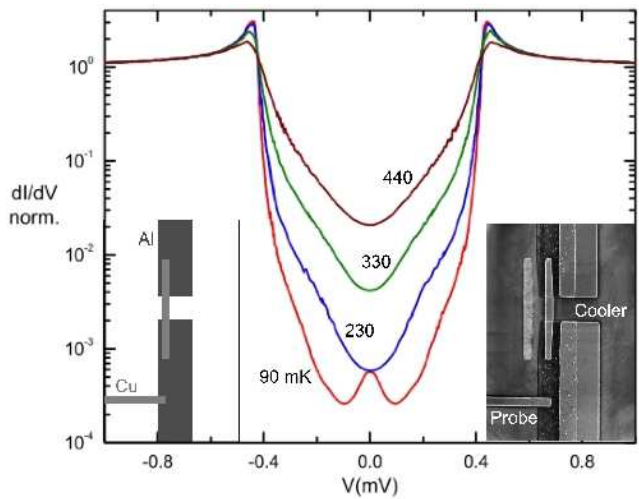}
\caption{(Color online) Left inset: Geometry of the sample. Right inset: Scanning electron microscope micrograph of a typical cooler sample made of a normal metal Cu electrode (light grey) connected to two superconducting Al reservoirs (dark grey) through tunnel junctions. One of the additional probe junctions connected to one Al reservoir is visible at the bottom. Main figure: normalized differential conductance as a function of the voltage and at the cryostat temperatures of 90 (red curve), 230 (blue), 330 (green) and 440 mK (brown).}
\label{DiffCond}
\end{center}
\end{figure}

Fig. \ref{DiffCond} inset shows a typical sample which features a geometry similar to the one studied in Ref. \cite{PRL-Rajauria}. It consists of a 50 nm thick, 4 $\mu$m long and 0.3 $\mu$m wide normal metal Cu electrode embedded between two 40 nm thick and 1.5 $\mu$m wide superconducting Al electrodes. The tunnel barriers at the two symmetric junctions of dimensions 1.5 $\times$ 0.3 $\mu m^2$ were prepared by oxidization in 0.2 mbar of O$_{2}$ pressure for 3 min and give a total normal-state resistance $R_n$ in the range 2-3 k$\Omega$. In addition to the two cooling junctions, we added three Cu tunnel probes of area 0.33$\times$0.43 $\mu m^2$ on one Al electrode (one is shown in Fig. \ref{DiffCond} insets). These probes are strongly connected to both a Cu reservoir and an Al reservoir so that no cooling is expected there. These junctions thus provide a reference for an isotherm behavior. In the following, we will describe the behavior of one out of three investigated samples which all showed similar behavior.

We have measured the current-voltage $I(V)$ characteristic of every probe junction and of the two cooling junctions in series at temperatures down to 90 mK. The differential conductance $dI/dV(V)$ is obtained by numerical differentiation. We have taken special care to obtain accurately the sub-gap conductance of our current-biased samples down to a level of about $10^{-4}$ of the normal state conductance. Fig. \ref{DiffCond}  displays on a logarithmic scale the differential conductance of the cooling junctions for a series of cryostat temperatures. The tunnel current in a N-I-S junction is given by:
\begin{equation}
I(V)=\frac{1}{eR_{n}}\int^{\infty}_{0}n_{s}(E)[f_{N}(E-\frac{eV}{2})-f_{N}(E+\frac{eV}{2})]dE,
\label{IT}
\end{equation}
where $V/2$ is the voltage across the N-I-S junction, $f_{N}(E)$ is the electron energy distribution function in the normal metal and $n_{s}(E)=\mid E \mid /\sqrt{E^2-\Delta^2}$ is the normalized BCS density of states in the superconductor. An isotherm N-I-S junction would then feature in the sub-gap regime a linear behavior of the differential conductance on a logarithmic scale. In contrast, we observe in the high temperature ($T >$ 200 mK) data an upward curvature which constitutes a clear signature of the electron cooling \cite{PRL-Rajauria}.

In the low temperature regime ($T <$ 200 mK), clearly a different characteristic is obtained with a differential conductance peak at zero bias. A similar behavior is obtained on every probe (not shown). This zero-bias peak cannot be accounted for by a single-particle tunnel current. The zero-bias differential conductance increases while the temperature is lowered below about 200 mK, which suggests that it is a phase-coherent effect.

\begin{figure}[t]
\begin{center}
\includegraphics[width=8.5 cm]{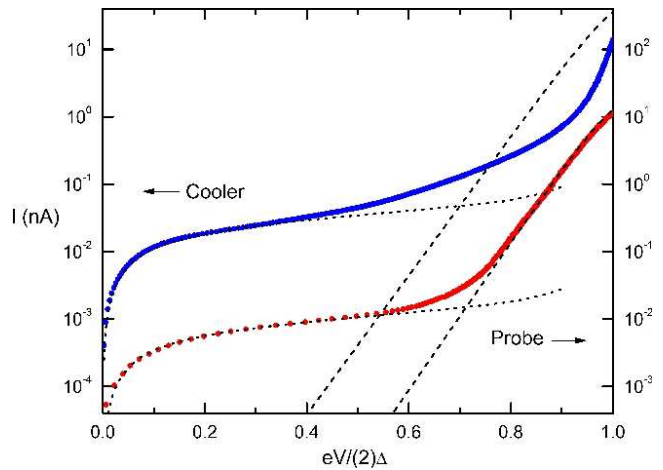}
\caption{(Color online) Current-voltage characteristic of the cooler junctions (blue symbols) and of the probe (red symbols) as a function of the voltage and at a cryostat temperature of 90 mK together with best-fit calculations of the Andreev current (dotted lines) and of the single-particle current (dashed lines) using the parameters $D=$ 80 $cm^2.s^{-1}$, $L_\varphi=$ 1.5 $\mu m$, $R_n$ = 1.9 $k\Omega$ (cooler) and 2.76 $k\Omega$ (probe), 2$\Delta$ = 0.43 meV (cooler) and $\Delta$ = 0.228 meV (probe). The fit temperature is 105 mK. Compared to the theoretical calculations based on Ref. \onlinecite{PRL-Hekking}, the Andreev current was multiplied by 1.55 (cooler) and 4.3 (probe). No cooling effect is included in the calculation.} \label{IV}
\end{center}
\end{figure}

We will ascribe the low bias differential conductance peak to an Andreev current, i.e. a double particle tunnel current created by Andreev reflections at the N-I-S junctions.  In order to calculate the Andreev current $I_A$, we used the theory of Ref. \cite{PRL-Hekking}. We took into account the finite gap $\Delta$ and the disorder both in the normal metal and in the superconductor. We considered the 1D regime where the coherence length of an Andreev pair in the normal metal $L_{E}=\sqrt{\hbar D/E}$ \cite{Superlattices-Courtois} is much larger than the junction dimension.

Let us first discuss the current-voltage characteristics of one probe junction, where the electronic temperature can be considered as constant and very close to the cryostat temperature. Fig. \ref{IV} displays the measured current-voltage characteristic of one probe junction (1.55 $\mu$m away from the cooler junctions) together with the calculated single quasi-particle and Andreev currents. We have fitted the current-voltage characteristics of the probe while taking the electron diffusion coefficient in Cu equal to the measured value $D=$ 80 $cm^2.s^{-1}$. The fit parameters are $\Delta$ = 0.228 meV, the electronic temperature $T_e =$ 105 mK, which is found slightly higher than the bath temperature, and $L_\varphi=$ 1.5 $\mu m$, which agrees well with expectations for a pure metal at very low temperature \cite{PhysicaScripta-Pannetier-PRL-Huard}. In this fit, we had to scale the Andreev current by a multiplying factor $M$= 4.3 in a similar way to Ref. \cite{PRL-Pothier}. The fit describes very well the probe data. The cross-over between the low-bias Andreev current and the high-bias single quasi-particle current is clearly visible. We did not need to take into account the possible contribution of pinholes in the barrier, which means that our junctions can be considered as ideal. We conclude that the sub-gap current in our N-I-S junctions is the superposition of a single-particle tunnel current and a two-particle Andreev current.

We tried to fit the S-I-N-I-S cooler junction characteristic in a similar way. Fig. \ref{IV} also shows the current-voltage characteristic of the cooler junction together with the calculated single quasi-particle and Andreev currents. A good fit to the theory is found only at small bias, where cooling or heating effect remain negligible. This fit is obtained with the same parameter set than for the probe, except for the gap 2$\Delta$ = 0.43 meV and the scaling factor $M =$ 1.55. The difference in the $M$ factor is presumably due to difference in geometry between the types of junctions.

\begin{figure}[t]
\begin{center}
\includegraphics[width=8.5 cm]{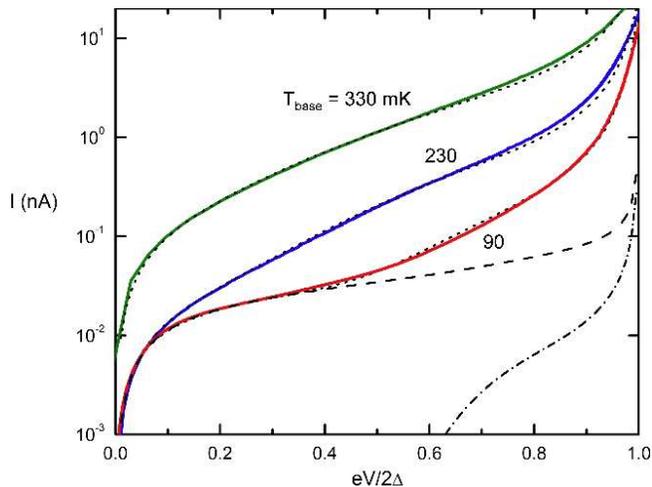}
\caption{(Color online) Current-voltage characteristic of the S-I-N-I-S cooler junction as a function the voltage at cryostat temperature of 90 (red line), 230 (blue) and 330 mK (green) together with best-fit calculated curves. Dot-dashed line: model including the cooling by the tunnel current, with parameters 2$\Delta$ = 0.43 meV, K.A = 144 $W.K^{-4}$. Dashed line: model including in addition the Andreev current, but not the related heating, with the parameters $D=$ 80 $cm^2.s^{-1}$, $L_\varphi=$1.5 $\mu m, M =$ 1.55. Dotted lines: full model taking into account the Andreev current and the related heating,} \label{SeriesOfT}
\end{center}
\end{figure}

In order to analyze quantitatively the behavior of our S-I-N-I-S cooler junctions, we need to consider the heat balance in the normal metal. Here we assume a quasi-equilibrium situation: the electrons and the phonons follow a thermal energy distribution functions at a respective temperature $T_e$ and $T_{ph}$ which are in general different from the cryostat temperature $T_{base}$. This is justified because the inelastic scattering time is of the order of the phase-coherence
time of 300 ps (derived from Fig. 2 fit), which is much shorter than the mean escape time from the island estimated to about 100 ns. The tunnel current is responsible for a cooling power in the normal metal:
\begin{equation}
P_{cool}=\frac{1}{e^2R_{n}}\int^{\infty}_{-\infty}(E-\frac{eV}{2})n_{s}(E)[f_{N}(E-\frac{eV}{2})-f_{S}(E)]dE
\end{equation}
where $f_{S}$ is the energy distribution function in the superconductor. This cooling power is compensated by the electron-phonon coupling power so that: $2P_{cool}+P_{e-ph}=0$. Here the factor 2 accounts for the presence of two junctions. We used the usual expression for the electron-phonon coupling $P_{e-ph}(T_{e},T_{ph})=\Sigma U(T_{e}^{5} - T_{ph}^{5})$, where $\Sigma$ is a material-dependent constant and $U$ is the metal volume. If one considers now the normal metal phonons, the electron-phonon coupling power is compensated by the Kapitza power  $P_{K}(T_{ph},T_{base})= KA(T_{base}^{4}-T_{ph}^{4})$, where $K$ is an interface-dependant parameter and $A$ the contact area, so that $P_{e-ph}+P_{K}=0$. The related Kapitza thermal resistance is significant only at intermediate temperature and above ($T>$ 300 mK), in which case the normal metal phonons can be cooled below the substrate temperature \cite{PRL-Rajauria}.

Fig. \ref{SeriesOfT} displays a series of current-voltage characteristics obtained at different temperatures together with calculated curves. We have first numerically solved the heat balance equations in the high temperature regime where the Andreev current contribution can be neglected. We took the well-accepted value $\Sigma$ = 2 nW.$\mu$m$^{-3}$.K$^{-5}$ \cite{JLTP-Godfrin} and we determined from the fitting procedure the Kapitza coupling parameter value $K.A =$ 144 $W.K^{-4}$.

Fig. \ref{SeriesOfT} dash-dotted line displays the calculated current-voltage characteristic calculated at a 90 mK cryostat temperature by including the charge and heat currents of only the single-particle channel effects. The agreement is poor, which confirms the need for including the Andreev current channel. As a second step, we included the Andreev charge current component. The dashed line shows the result of the calculation with again the same parameters set than in Fig. \ref{IV} and a cryostat temperature of 90 mK. It provides an acceptable fit at low bias but shows a clear discrepancy at intermediate voltage. Adding a leakage term in the heat balance with a linear resistance does not provide a good description of the data. Thus a significant thermal contribution is missing in the heat balance equations described above.

We investigated theoretically the heat transfer created by the Andreev current $I_A$~\cite{PRL-Chandrasekhar-Virtanen} flowing through an N-I-S tunnel junction~\cite{unpublished}, straightforwardly extending~\cite{PRL-Hekking}. We have found that the work performed by the current source feeding the circuit with a current $I_{A}$ generates a Joule heating $I_{A}V$ that is deposited in the normal metal. The full heat balance equation for the normal metal electrons can then be written as:
\begin{equation}
2P_{cool}+P_{e-ph}+I_{A}V=0.
\end{equation}
With this complete heat balance equation taken into account, we calculated the current-voltage characteristic at every cryostat temperature, see the dotted lines fits in Fig. \ref{SeriesOfT}. The agreement is very good for every accessible voltage and temperature. Assuming a thermalization of the phonons to the substrate temperature would change the current by less than 2$\%$ at 90 mK, which means that the phonon cooling has a negligible role in the data analysis presented here. Let us note that at the 90 mK cryostat temperature, the agreement covers 4 orders of magnitude for the current. 
\begin{figure}[t]
\begin{center}
\includegraphics[width=8.5 cm]{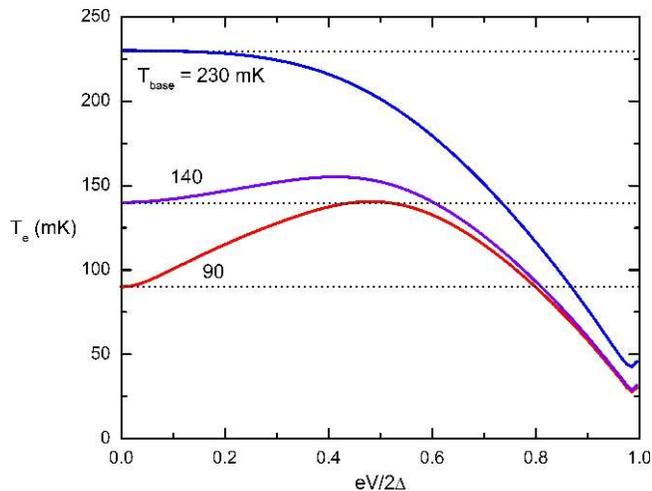}
\caption{(Color online) Dependence of the calculated electronic temperature with the voltage as derived from the fit of the experimental result and for a series of cryostat temperature: 90 (red symbols), 140 (green) and 230 mK (blue).} \label{Temperatures}
\end{center}
\end{figure}

This excellent agreement demonstrates that the heat balance in our S-I-N-I-S junctions can be fully understood by taking into account the contributions of both the single quasi-particle tunneling and the two-particle Andreev current. The numerical solution of the heat balance equations also provides the electron temperature at every bias. Fig. \ref{Temperatures} shows the calculated dependence of the normal metal central island electron temperature with the voltage across the cooler. At very low temperature, the electron temperature first increases with the bias due to Andreev current heating, before decreasing due to the tunnel current-based cooling effect. At a cryostat temperature of 90 mK, the cooling effect overcomes the Andreev current thermal contribution only close to the optimum bias. This demonstrates the great importance of the Andreev current-induced dissipation in a N-I-S tunnel junction. Although the Andreev current is a small effect in a such junction if one considers the charge current, this is no longer true if one considers the heat current. The explanation is the following. Compared to a Joule power $I.V$, the Andreev current contributes fully to heating while the tunnel current cools with a moderate efficiency \cite{RMP-Giazotto}. This efficiency is of the order of $T_e/\Delta$ which is about 5 $\%$ at a 100 mK electron temperature.

In conclusion, we have devised a quantitative analysis of the current-voltage characteristics of S-I-N-I-S hybrid junctions at very low temperature. Our study demonstrates the importance in terms of heat transport of the Andreev current which arises from the confinement of phase-coherent quasi-particles in the vicinity of the interface. This Andreev current-induced dissipation is with no doubt also of great importance in the behavior of S-N-S junctions with transparent interfaces, where a significant electron heating can be observed \cite{PRB-Hoss}.

The authors are grateful to Nanofab-CNRS for sample fabrication. We acknowledge useful discussions with J. P. Pekola, M. Houzet and A. Vasenko. This work was funded by EU STREP project 'SFINx' and by NanoSciERA project 'Nanofridge'.

 \end{document}